# Cascaded quantum time transfer breaking the no-cloning barrier with entanglement relay architecture

H. Hong[1,2], X. Xiang[1,2], R. Quan[1,2*], B. Shi[1,2], Y. Liu[1,2], Z. Xia[1,2], T. Liu[1,2,4], X. Li[1,2], M. Cao[1,2], S. Zhang[1,2,4], K. Guo[3*], R. Dong[1,2,4*]

[1] *Key Laboratory of Time Reference and Applications, National Time Service Center, Chinese Academy of Sciences, Xi'an 710600, China*
[2] *School of Astronomy and Space Science, University of Chinese Academy of Sciences, Beijing, 100049, China*
[3] *Institute of Systems Engineering, Academy of Military Science, Beijing, 100141, China*
[4] *Quantum Precision Measurement Department, Hefei National Laboratory, Hefei, 230026, China*

Quantum two-way time transfer (Q-TWTT) leveraging energy-time entangled biphotons has achieved sub-picosecond stability but faces fundamental distance limitations due to the no-cloning theorem's restriction on quantum amplification. To overcome this challenge, we propose a cascaded Q-TWTT architecture employing relay stations that generate and distribute new energy-time entangled biphotons after each transmission segment. Theoretical modeling reveals sublinear standard deviation growth (merely $\sqrt{N}\times$ increase for $N\times$ equidistant segments), enabling preservation of sub-picosecond stability over extended distances. We experimentally validate this approach using a three-station cascaded configuration over 2×100 km fiber segments, demonstrating strong agreement with theory. Utilizing independent Rb clocks at end and relay stations with online frequency skew correction, we achieve time stabilities of 3.82 ps at 10 s and 0.39 ps at 5120 s. The consistency in long-term stability between cascaded and single-segment configurations confirms high-precision preservation across modular quantum networks. This work establishes a framework for long-distance quantum time transfer that surpasses the no-cloning barrier, providing a foundation for future quantum-network timing infrastructure.

## 1 Introduction

Precision time synchronization has evolved into a cornerstone technology for contemporary infrastructure, supporting various critical applications such as relativistic geodesy [1, 2], coherent distributed aperture radio telescopes [3, 4], and quantum network synchronization [5]. The two-way time transfer (TWTT) technique [6] offers significant advantages in practical time synchronization applications by canceling out the substantial impacts of link delays and their associated fluctuations, thereby removing the need for intricate feedback compensation systems. Security has also become a vital concern, as malicious spoofing attacks could disrupt critical infrastructure reliant on precise timing [7, 8]. Notably, the TWTT has been proven effective in identifying man-in-the-middle (MITM) delay attacks [9, 10], further underscoring its operational value.

Fiber-based TWTT systems [11-14] leverage the low-loss and high reliability of optical fibers to achieve superior time and frequency synchronization compared to satellite-based methods [15-18]. Nevertheless, even fiber-based TWTT, as well as its space-based counterpart, faces inherent precision limitations at the picosecond level due to classical noise constraints [19-21].

---

*Corresponding author (email: dongruifang@ntsc.ac.cn; quanrunai@ntsc.ac.cn; guokai07203@hotmail.com).

To overcome these classical limitations, quantum time transfer protocols have been proposed and validated in experimental demonstrations [22-32]. Notably, the quantum TWTT (Q-TWTT) protocol [29-32] employs the stringent temporal correlation of energy-time entangled biphotons to enable precise determination of the time difference between distributed photons. By integrating the TWTT mechanism with quantum entanglement, this approach offers not only quantum-enhanced precision [30, 33] but also inherent security [34], representing a transformative advancement in synchronization technology. Nonetheless, due to the quantum no-cloning theorem, the optical amplification commonly used in conventional time transfer systems for extending fiber distances cannot be applied to the Q-TWTT scheme, posing challenges for its application in long-haul fiber-link transfers.

In this paper, we propose a cascaded quantum time-transfer (Q-TWTT) architecture deploying relay stations that generate and distribute new energy-time entangled biphotons for each transmission segment. Through theoretical analysis, we demonstrate that the standard deviations (SDs) of the achieved time offsets follow a sublinear SD growth—scaling only as $\sqrt{N}$ for $N \times$ equidistant segments. This stands in stark contrast to the exponential increase that would occur without the cascaded architecture, thus enabling the preservation of sub-picosecond time stability over extended distances. To validate the approach, we implement a three-station cascaded Q-TWTT experiment over 2×100 km fiber segments. Time transfer performance was evaluated under three clock reference configurations: (1) common reference clock (CRC) where a Rb clock was shared across the three stations; (2) independent reference clock (IRC) with two independent Rb clocks applied for the end stations and the relay station respectively; and (3) IRC with frequency skew correction (IRC-FC). The measured time offset SDs in all cases demonstrate close agreement with theoretical simulations. Time deviation (TDEV) analysis further reveals the IRC-FC stability of 3.82 ps at 10 s and 0.39 ps at 5120 s. This performance shows excellent consistency with the CRC results, confirming the efficacy of the online frequency skew correction. Critically, the alignment in long-term TDEV between the IRC-FC cascaded system and single-segment Q-TWTT demonstrates that the cascaded architecture preserves quantum-enhanced time stability across extended distances. This achievement establishes a long-distance quantum time transfer framework that surpasses the no-cloning barrier, while simultaneously providing the foundational building blocks for future quantum-network-compatible timing infrastructure. It enables entanglement-secured synchronization critical for enhancing reliability in fiber-based quantum communications and metrology networks.

## 2 Principle of the cascaded Q-TWTT scheme

Fig. 1 illustrates the proposed cascaded Q-TWTT architecture where relay stations are deployed in cascade after successive transmission, enabling periodic reestablishment of quantum entanglement distribution. For a system with M re-lay stations ($R_1, R_2, ... R_M$), the cascaded Q-TWTT system decomposes into M+1 single-segment Q-TWTT subsystems. Consequently, the total end-to-end time offset between A and B, denoted as $t_0 = t_{B,0} - t_{A,0}$, is the cumulative sum of time offsets between successive adjacent stations. Mathematically, this relationship can be expressed as:

$$t_0 = \sum_{k=0}^{M} \left( t_{k+1,0} - t_{k,0} \right), \quad (1)$$

where $t_{k,0}$ corresponds to the reference clock time at the $k$-th relay station $R_k$; $k=0$ and $k=M+1$ respectively correspond to the end stations A and B. This additive property also manifests the scalability of the cascaded quantum time transfer framework. As long as the $M+1$ single-segment Q-TWTT subsystems are mutually independent, the SD of $t_0$ is determined by

$$\Delta t_0 = \sqrt{\sum_{k=0}^{M} \Delta^2 \left( t_{k+1,0} - t_{k,0} \right)}. \quad (2)$$

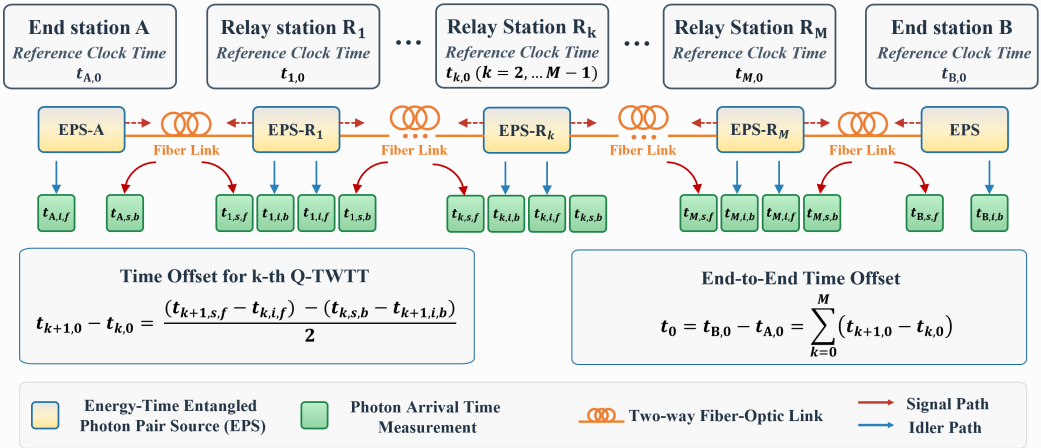

**Figure 1** System structure of the cascaded quantum two-way time transfer (Q-TWTT). The system aims to synchronize remote clocks located at two end stations A and B via dedicated fiber links with intermediate relay stations ($R_1, R_2, ...R_M$). Each relay station integrates an energy-time entangled photon pair source, quantum measurement modules, and a local reference clock. Entangled photon pairs are split into backward and forward propagation paths to perform Q-TWTT processes between adjacent stations (e.g., A↔$R_1$, $R_1$↔$R_2$, ..., $R_M$↔B). This cascaded design enables quantum-enhanced precision time transfer across extended distances in modular network configurations.

To perform the cascaded Q-TWTT procedure, every station is equipped with an energy-time entangled photon pair source, quantum measurement modules, and a local reference clock. At each relay station, the entangled photon pairs generated by the source are split into two separate parts for establishing bipartite entanglement links: one dedicated to conducting Q-TWTT with the preceding station (in the backward direction), and the other to performing Q-TWTT with the subsequent station (in the forward direction). The forward transmitted component of photon pairs at $R_k$ and the backward transmitted component at $R_{k+1}$ jointly support the $k$-th Q-TWTT procedure: at station $R_k$, the single photon detector $D_{k,i,f}$ monitors the locally retained idler photons of the forward transmitted component, while the detector $D_{k,s,b}$ registers the incoming signal photons from the backward transmitted component sent by station $R_{k+1}$. Similarly, at station $R_{k+1}$, the detector $D_{k+1,i,b}$ observes the locally retained idler photons in the backward transmitted component, and the detector $D_{k+1,s,f}$ receives the incoming signal photons from the forward transmitted component originating at station $R_k$. The recorded four series of timestamps are correspondingly denoted as $t_{k,i,f}$, $t_{k,s,b}$, $t_{k+1,i,b}$, $t_{k+1,s,f}$. Then, by performing the nonlocal coincidence identification on the entangled photon pairs [35], the bidirectional time difference terms, denoted by $(t_{k+1,s,f} - t_{k,i,f})$ and $(t_{k,s,b} - t_{k+1,i,b})$, are extracted. The relevant time offset $(t_{k+1,0} - t_{k,0})$ for single-segment Q-TWTT can be given by

$$t_{k+1,0} - t_{k,0} = \frac{(t_{k+1,s,f} - t_{k,i,f}) - (t_{k,s,b} - t_{k+1,i,b})}{2}. \quad (3)$$

Statistically, the SD of $t_{k+1,0} - t_{k,0}$ is determined by

$$\Delta(t_{k+1,0} - t_{k,0}) = \frac{1}{2}\sqrt{\Delta^2(t_{k+1,s,f} - t_{k,i,f}) + \Delta^2(t_{k,s,b} - t_{k+1,i,b})}, \quad (4)$$

where $\Delta(t_{k+1,s,f} - t_{k,i,f})$ and $\Delta(t_{k,s,b} - t_{k+1,i,b})$ denote the SDs of the two measured forward and backward transmission delays, respectively. According to the quantum theoretical model, the determination of every individual SD is associated with the relevant coincidence measurement. Thus, based on the detection events from $D_{k+1,s,f}$ and $D_{k,i,f}$ over a certain measurement interval, $\Delta(t_{k+1,s,f} - t_{k,i,f})$ can be determined by the measured two-photon temporal coincidence width $\sigma_{k,k+1,f}$, photon pair counts $P_{k,k+1,f}$ as well as the coincidence-to-accidental ratio $CAR_{k,k+1,f}$ [36], which can be given by[37]

$$\Delta(t_{k+1,s,f} - t_{k,i,f}) = \frac{\sigma_{k,k+1,f}}{\sqrt{2P_{k,k+1,f}/(1+1/CAR_{k,k+1,f})}}. \quad (5)$$

Equivalently, $\Delta(t_{k,s,b} - t_{k+1,i,b})$ can be expressed as

$$\Delta(t_{k,s,b} - t_{k+1,i,b}) = \frac{\sigma_{k,k+1,b}}{\sqrt{2P_{k,k+1,b}/(1+1/CAR_{k,k+1,b})}}, \quad (6)$$

where $\sigma_{k,k+1,b}$, $P_{k,k+1,b}$, $CAR_{k,k+1,b}$ respectively represent the measured two-photon temporal coincidence width, photon pair counts as well as the CAR depending on the detection events from $D_{k,s,b}$ and $D_{k+1,i,b}$ over the equal measurement interval. Eq. (4) is thus rewritten as:

$$\Delta(t_{k+1,0} - t_{k,0}) = \frac{1}{2}\sqrt{\frac{\sigma_{k,k+1,f}^2}{2P_{k,k+1,f}/(1+1/CAR_{k,k+1,f})} + \frac{\sigma_{k,k+1,b}^2}{2P_{k,k+1,b}/(1+1/CAR_{k,k+1,b})}}. \quad (7)$$

Further substituting Eq. (7) into Eq. (2), the overall SD of the cascaded Q-TWTT system can be given by

$$\Delta t_0 = \frac{1}{2}\sqrt{\sum_{k=0}^{M}\frac{\sigma_{k,k+1,f}^2}{2P_{k,k+1,f}/(1+1/CAR_{k,k+1,f})} + \sum_{k=0}^{M}\frac{\sigma_{k,k+1,b}^2}{2P_{k,k+1,b}/(1+1/CAR_{k,k+1,b})}}. \quad (8)$$

To quantify the parameters in Eq. (8), assume the entangled photon pairs for both the forward and backward transmissions at station $R_k$ are generated with a photon number of $P_{0,k}$ within a measurement interval $T$. The link between stations $R_k$ and $R_{k+1}$ is fiber segment with a distance of $l_{k,k+1}$ and a propagation loss of $e^{-\alpha l_{k,k+1}}$, where $\alpha$ denotes the attenuation factor of the single mode fiber. While the quantum efficiencies of single photon detectors (SPDs) and other fixed system parameters introduce a constant loss of $\eta_k$. Consider the accidental coincidence counts within $T$ for each pair of SPDs are identical and amounts to $P_{ac}$. Thus,

$$P_{k,k+1,f} = P_{0,k}\eta_k e^{-\alpha l_{k,k+1}} + P_{ac}, \quad (9a)$$

$$P_{k,k+1,b} = P_{0,k+1}\eta_{k+1} e^{-\alpha l_{k,k+1}} + P_{ac}, \quad (9b)$$

$$CAR_{k,k+1,f} = \frac{P_{0,k}\eta_k e^{-\alpha l_{k,k+1}}}{P_{ac}}, \quad (9c)$$

$$CAR_{k,k+1,b} = \frac{P_{0,k+1}\eta_{k+1} e^{-\alpha l_{k,k+1}}}{P_{ac}}. \quad (9d)$$

Substituting these parameters into Eq. (8),

$$\Delta t_0 = \frac{1}{2}\sqrt{\sum_{k=0}^{M}\left(\frac{\sigma_{k,k+1,f}^2}{2P_{0,k}\eta_k e^{-\alpha l_{k,k+1}}} + \frac{\sigma_{k,k+1,b}^2}{2P_{0,k+1}\eta_{k+1} e^{-\alpha l_{k,k+1}}}\right)}. \quad (10)$$

The measured two-photon temporal coincidence width $\sigma_{k,k+1,f(b)}$ consists of three contributions: first, the correlation widths of the photon pairs after dispersive broadening in the fiber link between station $R_k$ and $R_{k+1}$, denoted as $\sigma_{k,f}$ for the forward-transmission component at $R_k$ and $\sigma_{k+1,b}$ for the backward-transmission component at $R_{k+1}$; second, the timing jitter of the SPDs for coincidence measurements, simplified as $\sigma_{jit}$ for all SPDs; and third, the frequency skew $\Delta u_{k,k+1}$ between the local reference clocks at $R_k$ and $R_{k+1}$, which contributes $\sigma_{clk,k,k+1} = \Delta u_{k,k+1} T$ over the measurement interval $T$ [38]. Thus $\sigma_{k,k+1,f(b)}$ can be rewritten as

$$\sigma_{k,k+1,f} = \sqrt{\sigma_{k,f}^2 + \sigma_{jit}^2 + \sigma_{clk,k,k+1}^2}, \quad (11)$$

$$\sigma_{k,k+1,b} = \sqrt{\sigma_{k+1,b}^2 + \sigma_{jit}^2 + \sigma_{clk,k,k+1}^2}. \quad (12)$$

With optimized nonlocal dispersion cancellation (NDC) [39,40], the contributions of $\sigma_{k,f}$ and $\sigma_{k+1,b}$ in Eqs. (11) and (12) become negligible. Assuming identical frequency skew between each pair of adjacent stations ($\Delta u_{k,k+1} \approx \Delta u$), the temporal coincidence widths can be deduced as $\sigma_{k,k+1,f} \approx \sigma_{k,k+1,b} \approx \sigma = \sqrt{\sigma_{jit}^2 + (\Delta u T)^2}$. Considering uniform parameters across all stations—identical photon-pair generation rate ($P_{0,k} = P_0$), equidistant fiber segments ($l_{k,k+1} = l$) and equal loss factors ($\eta_k = \eta$)—the overall SD $\Delta t_0(Nl)$ of the cascaded Q-TWTT system can be expressed as:

$$\Delta t_0(Nl) = \frac{1}{2}\sqrt{\sum_{k=0}^{M}\left(\frac{\sigma^2}{2P_0\eta e^{-\alpha l}} + \frac{\sigma^2}{2P_0\eta e^{-\alpha l}}\right)} = \frac{\sqrt{N}\sigma}{2\sqrt{P_0\eta e^{-\alpha l}}},$$

(13)

where $N = M+1$ representing the $N$ equidistant single-segment Q-TWTT subsystems of the cascaded Q-TWTT system. The SD for a single-segment Q-TWTT subsystem over distance $l$ is given by

$$\Delta t_{0,\sin}(l) = \frac{\sigma}{2\sqrt{P_0\eta e^{-\alpha l}}}. \quad (14)$$

By comparing Eqs. (13) and (14), for the cascaded Q-TWTT system over extended distance $Nl$, the SD is merely increased by

$$\frac{\Delta t_0(Nl)}{\Delta t_{0,\sin}(l)} = \sqrt{N}. \quad (15)$$

Critically, without the cascaded architecture, the SD of Q-TWTT system over distance $Nl$ would become

$$\Delta t_{0,\sin}(Nl) = \frac{\sigma}{2\sqrt{P_0\eta e^{-N\alpha l}}}. \quad (16)$$

This non-cascaded SD grows exponentially as $e^{N\alpha l/2}$, significantly exceeding the $\sqrt{N}$ scaling. The cascaded architecture thus efficiently mitigates fiber loss-induced SD accumulation, enabling preservation of sub-picosecond time stability over extended distances.

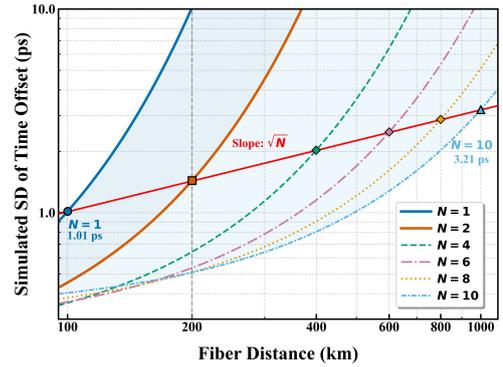

**Figure 2** Simulated $\Delta t_0(Nl)$ versus fiber propagation distance for cascaded Q-TWTT systems with different numbers of single-segment Q-TWTT subsystems ($N$=1-10). The blue solid curve represents the system without cascaded architecture ($N$=1), while other lines indicate systems with increasing numbers of single-segment Q-TWTT from $N$=2 (orange solid curve) to $N$=10 (blue dash-dotted curve). The red solid curve represents the theoretical simulation of the attainable SDs with equidistant 100 km fiber segments. All simulations account for experimental parameters including entangled photon pair generation rates of 2.5×10$^6$ pairs per measurement interval, 11 dB loss for idler photons, 0.2 dB/km fiber propagation loss for signal photons, and 64 ps SPD timing jitter.

Using our experimental parameters, we modeled entangled photon pair sources emitting 2.5×10$^6$ pairs per measurement interval for both forward and backward transmissions. The idler photons undergo a total loss of 11 dB, comprising insertion losses from fiber components (FPBS, FBS, FBG) and limited quantum efficiencies of SPDs. For signal photons, they experience a propagation loss of 0.2 dB/km in the fibers plus 3 dB loss from the involved fiber compo-nents. In the absence of frequency skew and with the temporal coincidence width dominated by 64 ps SPD timing jitter, Fig. 2 shows the simulated $\Delta t_0(Nl)$ versus fiber propagation distance for systems with an increasing number of equidistant Q-TWTT segments.

As shown in Fig. 2, for non-cascaded Q-TWTT systems with 100-km fiber segments between adjacent stations, the SD is 1.01 ps. When cascading 9 relay stations across successive 100-km segments (10 equidistant Q-TWTT segments with total distance of 1000 km), the system SD increases by only $\sqrt{10}$ to 3.21 ps—consistent with the expected $\sqrt{N}$ scaling law. It is anticipated that the system SD can be further reduced by shortening the segment length while increasing the number of relay stations. For instance, with 50-km fiber segments, the non-cascaded Q-TWTT SD decreases to 0.32 ps, and extending to 1000 km via 19 relay stations then merely increases the overall SD to 1.43 ps.

## 3  Experimental setup

Subsequently, experimental validation was performed using a three-station cascaded configuration over 2×100 km fiber segments. As illustrated in Fig. 3, end stations A and B connect to relay station R via separate 100-km fiber segments. All three stations are situated within a single laboratory. Each station is equipped with a custom-built, all-fiber energy-time entangled photon pair source (EPS) [41] operating in the telecom band, designated as EPS-A (station A), EPS-R (relay station R), and EPS-B (station B), respectively. Each EPS is coupled to a fiber-based polarization beamsplitter (FPBS) that separates the polarization-orthogonal photon pairs into two distinct propagation paths. At the relay station R, the photon pairs (EPS-R) are further divided into two components: one forward entanglement distributing (EPS-R$f$) and the other backward entanglement distributing (EPS-R$b$). This configuration enables the sequential Q-TWTT procedures between A↔R and R↔B through bidirectional transmission of signal photons from EPS-A/B and EPS-R$f$/$b$ to their counterpart stations. The transmission is facilitated by the fiber links and optical circulators (OC1&OC2 for A↔R, OC3&OC4 for R↔B), while corresponding idler photons remain at their local stations for coincidence measurements.

To nonlocally compensate for the dispersions across the 100 km fiber links between stations A/B and station R, each idler photon path incorporates a dedicated dispersion-compensating fiber Bragg grating (FBG, Proximion Inc). These FBGs follow a systematic labeling convention based on their deployment locations and propagation directions: FBG-A (station A), FBG-R$b$ (backward-distributing at station R), FBG-R$f$ (forward-distributing at station R), and FBG-B (station B), ensuring unambiguous identification of their roles within the optical network. After the FBGs, the idler photons are routed through variable optical attenuators (VOAs, to avoid the saturation of the single photon detectors, not shown in the figure,) and then combined with incoming signal photons from neighboring stations using dedicated fiber beamsplitters. Following the same labeling convention as the FBGs, these beamsplitters are labeled as FBS-A, FBS-R$b$, FBS-R$f$, and FBS-B, respectively. The combined photon streams from each FBS are sent to four superconducting nanowire single-photon detectors (SNSPDs, Photon Technology Co., Ltd., labeled D1-D4), each with a quantum efficiency of ~80% and a timing jitter of ~64 ps. To prevent saturation, the photon count rates were settled at ~200 kcps by adjusting the aforementioned VOAs. The electrical outputs from the SNSPDs were time tagged by station-specific time-tagging units (TTUs, Swabian Instruments)—TTU-A, TTU-R, and TTU-B—each referenced to a dedicated high-stability atomic clock (labeled Clock-A, Clock-R, and Clock-B). Ad-ditionally, a programmable phase trimmer (PPT, Synchronization Technology Ltd) was connected between clock-R and TTU-R to execute the frequency skew correction based on the time offset measurement between stations A and R.

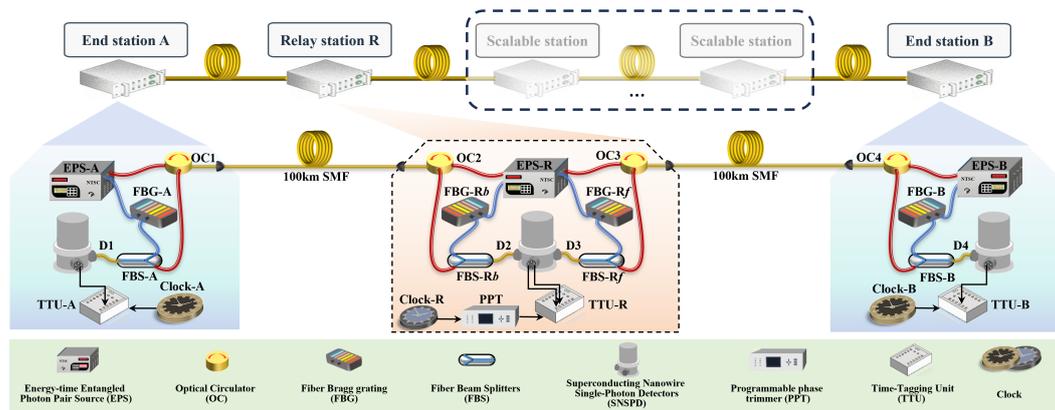

**Figure 3** Experimental diagram of the three-station cascaded Q-TWTT system with stations A and B connected through relay station R via two 100 km fiber links. Each station is equipped with an energy-time entangled photon pair source (EPS), fiber Bragg gratings (FBGs) for nonlocal dispersion compensation, superconducting nanowire single-photon detectors (SNSPDs) for photon detection, fiber beamsplitters (FBS), and time-tagging units (TTUs) referenced to atomic clocks. Optical circulators (OC1-OC4) facilitate bidirectional transmission between stations. A programmable phase trimmer (PPT) is connected between Clock-R and TTU-R to perform frequency skew correction based on the measured time offset between stations A and R.

## 4  Results and analysis

To assess the cascaded Q-TWTT system performance, we implemented three distinct clock configurations: (1) common reference clock (CRC) where all three stations shared a common Rb clock (SRS Inc.), yielding a baseline time offset $t_{0,CRC}$; (2) independent reference clock (IRC) with an independent Rb clock (Rock Electronic Ltd.) utilized as Clock-R, resulting in the time offset $t_{0,IRC}$; (3) IRC with online frequency skew correction via the PPT connected between Clock-R and TTU-R, delivering the time offset $t_{0,IRC-FC}$.

In the CRC configuration, the temporal coincidence widths $\sigma_{A,R,f}$, $\sigma_{A,R,b}$, $\sigma_{R,B,f}$, $\sigma_{R,B,b}$ were measured to be 74.90±2.52 ps, 73.66±3.08 ps, 124.74±19.20 ps, and 119.82±5.68 ps, respectively. The inconsistency between these measured widths is mainly induced by imperfect NDC of the fiber links, especially for R↔B link. The correspondingly measured photon pair counts for $P_{A,R,f}$, $P_{A,R,b}$, $P_{R,B,f}$, $P_{R,B,b}$ were approximately 1258±79, 857±41, 373±73, and 940±89, respectively within the measurement interval of 10 s. The CARs were measured to be around 31.3, 22.6, 2.8, and 8.8, respectively. Using Eq. (7), we calculated the time offset SDs to be 1.18 ps (for $t_{0,CRC}^{A↔R}$), 3.00 ps (for $t_{0,CRC}^{R↔B}$), and 3.22 ps (for $t_{0,CRC}$), with these values serving as the initial points (frequency skew $\Delta u$ =0) for the orange-square, green-triangle, and blue-circle curves in Fig. 4, respectively. The measured $\Delta t_{0,CRC}$ was determined to be 3.54 ps and is marked by the blue star in Fig. 4. It can be seen that, good agreement is reached between the measurement and the simulation.

To model the frequency skew dynamics in the IRC configuration, we simulated $\Delta t_{0,IRC}^{A↔R}$, $\Delta t_{0,IRC}^{R↔B}$, and $\Delta t_{0,IRC}$ as a function of the frequency skew ($\Delta u$) between stations A/B and R. As illustrated in Fig. 4 (orange-square curve for $\Delta t_{0,IRC}^{A↔R}$, green triangle curve for $\Delta t_{0,IRC}^{R↔B}$, and blue-circle curve for $\Delta t_{0,IRC}$), all simulated SDs exhibit a monotonic increase with rising $\Delta u$. For $\Delta u$ <5×10$^{-12}$, where frequency skew-induced temporal broadening of the coincidence widths remain negligible compared to those contributed from the transmission link and the detector jitter, $\Delta t_{0,IRC}$ grows only moderately compared to $\Delta t_{0,CRC}$.

When $\Delta u$ exceeds 5×10$^{-12}$, the frequency skew-induced coincidence width broadening and CAR degradation of the photon pairs become increasingly pronounced. At $\Delta u$ =1.90×10$^{-11}$, $\Delta t_{0,IRC}$ degrades by a factor of 2, reaching 6.44 ps. Further increasing the skew to 3.26×10$^{-11}$ pushes $\Delta t_{0,IRC}$ beyond 10 ps, surpassing the limit for high-precision time transfer and rendering sub-picosecond transfer infeasible without active frequency skew correction.

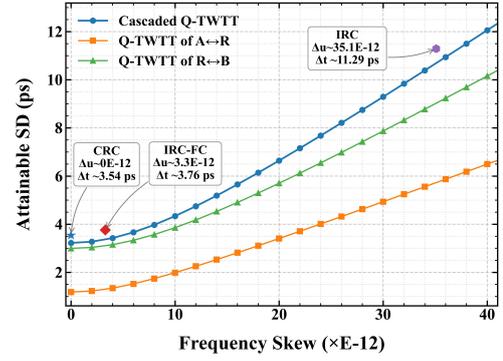

**Figure 4**  Simulated SDs of $t_{0,IRC}^{A↔R}$ (orange-square curve), $t_{0,IRC}^{R↔B}$ (green-triangle curve), and $t_{0,IRC}$ (blue-circle curve) as a function of frequency skew. The measured $\Delta t_{0,CRC}$, $\Delta t_{0,IRC}$, and $\Delta t_{0,IRC-FC}$ are also given by blue star, purple hexagon, and red diamond respectively for comparison with the simulations.

The temporal evolution of $t_{0,CRC}$, $t_{0,IRC}$, and $t_{0,IRC-FC}$ over the measurement duration of 35000 s are plotted in Fig. 5(a)-(c), with absolute time offsets removed for direct comparison. In the CRC configuration, $t_{0,CRC}$ behaves homogeneity, with its SD to be 3.54 ps (as shown in Fig. 4 by a blue star), which closely aligns with the simulated SD of 3.22 ps. As the composite $t_{0,CRC}$ is derived from the single-segment Q-TWTT measurements between A↔R and R↔B, the individual contributions $t_{0,CRC}^{A↔R}$ (orange curve) and $t_{0,CRC}^{R↔B}$ (green curve) are also shown in Fig. 5. The slight drifts appeared in these two curves arise from the residual frequency skew ($\Delta u_{R,PPT}$) of the equipped PPT in station R even if it is turned off. Via polynomial fitting [42], a value of $\Delta u_{R,PPT}$ ~8×10$^{-16}$ is yielded, which is much lower than 5×10$^{-12}$, resulting in negligible impact on $\Delta t_{0,CRC}$. The results demonstrate that the R↔B Q-TWTT exhibits significantly degraded SD compared to the A↔R Q-TWTT, primarily due to its lower photon pair count rate and suboptimal induced temporal broadening of the coincidence width. Consequently, the achievable SD of the cascaded Q-TWTT system under CRC configuration is fundamentally determined by the R↔B Q-TWTT performance.

In the IRC configuration, the frequency skew of the two independent Rb clocks was measured as $\Delta u_{IRC} = (35.1 \pm 0.01) \times 10^{-12}$, derived from the temporal evolutions of $t_{0,IRC}^{A \leftrightarrow R}$ and $t_{0,IRC}^{R \leftrightarrow B}$ [42]. Following the simulation shown in Fig. 4, $\Delta u_{IRC}$ would lead to deterioration of $\Delta t_{0,IRC}$ to a value of 10.69 ps. For comparison, $\Delta t_{0,IRC}$ was measured to be 11.29 ps (purple hexagon in Fig. 4). Its consistency with the simulation well confirms the significant skew-induced impact on the measured $\Delta t_{0,IRC}$.

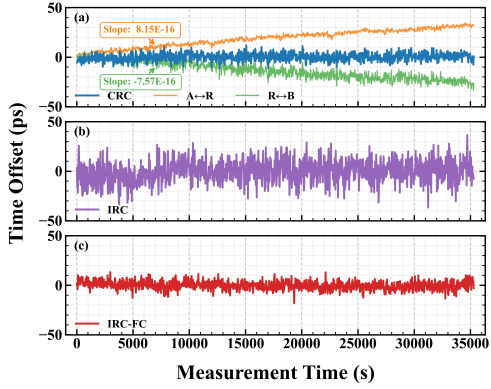

**Figure 5** Measured time offsets of the cascaded Q-TWTT system as a function of the measurement time under three scenarios: (a) CRC (blue curve), with the orange and green curves corresponding to the A↔R and R↔B link respectively; (b) IRC (purple curve), and (c) IRC-FC (red curve).

By turning on the PPT in station R and applying the feedback based on the measured $t_{0,IRC}^{A \leftrightarrow R}$, the significant frequency skew between the two independent Rb clocks was effectively suppressed to the level of $(3.3 \pm 0.02) \times 10^{-12}$—lower than $5 \times 10^{-12}$ shown in Fig. 4—resulting in slight influence on the SD of time offset. The temporal evolution of $t_{0,IRC-FC}$ is shown in Fig. 5 (c), which results in a SD value of 3.76 ps and is marked by a red diamond in Fig. 4. The nice agreement between theoretical and experimental results highlights the necessity of frequency skew correction for maintaining high performance in cascaded Q-TWTT systems.

The time stability performance in terms of TDEV in the CRC scenario is depicted in Fig. 6 by blue circles, with TDEV results of 3.64 ps at 10 s averaging time. For comparison, under CRC condition, the inset depicts the TDEVs of A↔R (orange squares) and R↔B (green triangles) single-segment Q-TWTTs. At averaging time of 10 s, the TDEV results are 1.10 ps and 3.46 ps for A↔R and R↔B, respectively. It is seen that the TDEV results in the CRC scenario maintain close alignment with that in R↔B Q-TWTT segment, just as the performance of SDs. With the averaging time beyond 2000 s, the TDEV results of the cascaded system align well with A↔R and R↔B Q-TWTT segments, and a minimum of 0.39 ps at 5120 s can be achieved. The consistency in long-term stability of the cascaded system and A↔R/R↔B Q-TWTT segments confirms high-precision preservation over extended distances by cascaded architecture.

Under the IRC condition (purple hexagons), the TDEV of 10.37 ps at 10-s averaging time corroborates the aforementioned SD measurement, and manifests the frequency skew-induced degradation on the achievable time stability. For IRC-FC case (red diamonds), the TDEV of 3.82 ps at 10 s shows a less than 0.2 ps deviation from the CRC result, demonstrating the effectiveness of our frequency skew-correction scheme. All the three TDEV curves (CRC, IRC and IRC-FC) decrease in a $\sqrt{\tau}$ trend within an averaging time of 1000 s. At averaging times around 2000 s, similar protruding bulges manifest the common circumstance disturbance of the laboratory [31]. At 5120 s, the TDEV of the IRC case evolves down to 0.88 ps. While for the IRC-FC case, the TDEV reaches a minimum of 0.39 ps, approaching the CRC performance level, further validating the necessity of frequency skew correction.

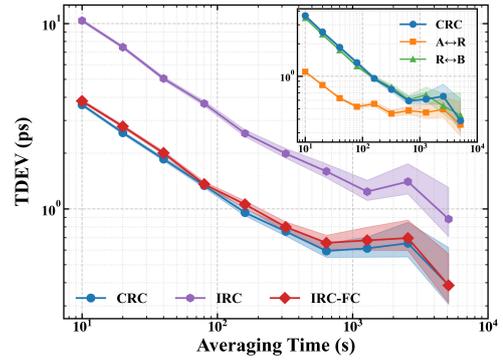

**Figure 6** Measured TDEVs of the cascaded Q-TWTT system under CRC (blue circles), IRC (purple hexagons), and IRC-FC (red diamonds) scenarios; the inset shows the CRC TDEVs for A↔R (orange squares) and R↔B (green triangles) Q-TWTT segments.

## 5 Discussion and conclusions

In summary, we have demonstrated a cascaded Q-TWTT architecture that overcomes fundamental distance limitations imposed by the no-cloning theorem. By deploying entanglement relay stations that regenerate entangled biphotons after each transmission segment, our model predicts sublinear growth in timing standard deviation with the increase in the number of equidistant Q-TWTT segments. Experimental validation using a three-station cascaded configuration over 2×100 km fiber segments, under three distinct clock reference configurations, yielded results in close agreement with theoretical simulations. Crucially, the sub-picosecond time stability (0.39 ps at 5120 s) observed under both CRC and IRC-FC conditions matches that of single-segment Q-TWTTs, confirming the architecture's capability to preserve quantum-enhanced timing stability over extended distances. These results establish a scalable pathway toward thousands-kilometers-scale quantum time transfer networks. Future performance improvements of at least one order of magnitude are attainable through enhanced entangled biphoton source brightness (e.g., GHz-rate sources [43]) and reduced two-photon temporal coincidence widths (e.g., via picosecond-jitter single-photon detectors [44]). The inherent compatibility of this quantum time transfer solution ensures its seamless integration into emerging quantum networks, providing the foundational timing infrastructure essential for advanced quantum communications [45-47] and precision metrology [48, 49].

*This work was supported by the National Natural Science Foundation of China (Grant Nos. 12033007, 12103058, 12203058, 12074309), Youth Innovation Promotion Association of the Chinese Academy of Sciences (2021408, 2022413, 2023425), and the Innovation Program for Quantum Science and Technology (2021ZD0300900).*

**Conflict of interest**   The authors declare that they have no conflict of interest.

1. J. Miller-Jones, A. Bahramian, J. Orosz, I. Mandel, L. Gou, T. Maccarone, C. Neijssel, X. Zhao, J. Ziółkowski, M. Reid, P. Uttley, X. Zheng, D.-Y. Byun, R. Dodson, V. Grinberg, T. Jung, J.-S. Kim, B. Marcote, S. Markoff, M. Rioja, A. Rushton, D. Russell, G. Sivakoff, A. Tetarenko, V. Tudose, J. Wilms, Science **371**, 1046–1049 (2021).
2. M. Bagherbandi, M. Shirazian, H. Amin, M. Horemuz, Front. Earth Sci. **11**, 1139211. (2023).
3. M. Xin, K. Shafak, and F. Kärtner, Optica **5**(12), 1564–1578 (2018).
4. M. Rioja, R, Dodson, Astron Astrophys Rev **28**, 6 (2020).
5. Y. Chen, Q. Zhang, T. Chen, W. Cai, S. Liao, J. Zhang, K. Chen, J. Yin, J. Ren, Z. Chen, S. Han, Q. Yu, K. Liang, F. Zhou, X. Yuan, M. Zhao, T. Wang, X. Jiang, L. Zhang, W. Liu, Y. Li, Q. Shen, Y. Cao, C. Lu, R. Shu, J. Wang, L. Li, N. Liu, F. Xu, X. Wang, C. Peng, J. Pan, Nature **589**, 7841, 214–219 (2021).
6. Ł. Śliwczyński, P. Krehlik, and M. Lipiński, Meas. Sci. Technol., **21**, 7:75302 (2010).
7. W. Gao, H. Li, M. Zhong and M. Lu, IEEE Trans. Smart Grid, **14**, 6:4784–4798 (2023).
8. J. Lee, E. Schmidt, N. Gatsis, and D. Akopian, IEEE Access, **11**, 138986–139003 (2023).
9. L. Narula and T. E. Humphreys, IEEE J. Sel. Topics Signal Process, **12**, 4, 749–762, (2018).
10. H. Dai, Q. Shen, C. Wang, S. Li, W. Liu, W. Cai, S. Liao, J. Ren, J. Yin, Y. Chen, Q. Zhang, F. Xu, C. Peng, J. Pan, Nat. Phys., **16**, 8, 848-852 (2020).
11. Ł. Śliwczyński, P. Krehlik, M. Lipiński H. Ender, H. Schnatz, D. Piester, and A. Bauch, IEEE Commun. Mag., **58**, 4, 67–73 (2020).
12. M Rost, D Piester, W Yang, T Feldmann, T Wübbena, and A Bauch, Metrologia, **49**, 772-778 (2012).
13. P. Krehlik, Ł. Sliwczynski, Ł. Buczek, M. Lipinski, IEEE Trans. Instrum. Meas. **61**, 2844-2851 (2012).
14. L. Wang, Y. Liu, W. Jiao, L. Hu, J. Chen, and G. Wu, Opt. Express **30**, 14, 25522-25535 (2022).
15. P. Defraigne, E. Pinat, and B. Bertrand, GPS Solutions, **25**, 2 45 (2021).
16. L. Liu, G. Tang, C. Han, X. Shi, R. Guo, L. Zhu, Sci. China-Phys. Mech. Astron. **58**, 89502 (2015).
17. S. Zhou, X. Hu, L. Liu, R. Guo, L. Zhu, Z. Chang, C. Tang, X. Gong, R. Li, Y. Yu, Sci. China-Phys. Mech. Astron. **59**, 109511 (2016).
18. W. Wang, X. Yang, S. Ding, W. Li, H. Su, P. Wei, F. Cao, L. Chen, J. Gong, Z. Li, IEEE Trans. Ultrason. Ferroelectr. Freq. Control, **65**, 8: 1475-1486 (2018).
19. E. Dierikx, A. Wallin, T. Fordell, J. Myyry, P. Koponen, M. Merimaa, T. Pinkert, J. Koelemeij, H. Peek and R. Smets, IEEE Trans. Ultrason. Ferroelectr. Freq. Control, **63**, 7, 945-952 (2016).
20. H. Zhang, G. Wu, L. Hu, X. Li, J. Chen, IEEE Photon. J. **7**, 6 7600208 (2015).
21. X. Guo, B. Hou, B. Liu, F. Yang, W. Kong, T. Liu, R. Dong, S. Zhang, Chin. Phys. Lett. **41**, 064202 (2024).
22. R. Jozsa, D. S. Abrams, J. P. Dowling, and C. P. Williams, Phys. Rev. Lett., **85**, 9, 2010 (2000).
23. V. Giovannetti, S. Lloyd, and L. Maccone, Nature, **412**, 417–419 (2001).
24. J. Zhang, G. Long, Z. Deng, W. Liu, and Z. Lu, Phys. Rev. A, **70**, 6, 62322 (2004).
25. J. Wang, Z. Tian, J. Jing, and H. Fan, Phys. Rev. D, **93**, 6, 65008 (2016).
26. A. Valencia, G. Scarcelli, and Y. Shih, Appl. Phys. Lett., **85**, 13, 2655–2657 (2004).
27. J. Lee, L. Shen, A. N. Utama, and C. Kurtsiefer, Opt. Express, **30**, 11, 18530–18538 (2022).
28. B. Tang, M. Tian, H. Chen, H. Han, H. Zhou, S. Li, B. Xu, R. Dong, B. Liu, W. Yu, EPJ Quantum Technol., **10**, 1 (2023).
29. F. Hou, R. Dong, T. Liu, and S. Zhang, in Proc. Quantum Inf. Meas., 2017, pp. QF3A–QF34 (2017).
30. F. Hou, R. Quan, R. Dong, X. Xiang, B. Li, T. Liu, X. Yang, H. Li, L. You, Z. Wang, S. Zhang, Phys. Rev. A, **100**, 023849 (2019).
31. H Hong, R Quan, X Xiang, Y Liu, T Liu, M Cao, R Dong, S Zhang, J. Lightw. Technol. **42**, 5, 1479-1486 (2024).
32. J. Lee, L. Shen, A. Cerè, J. Troupe, A. Lamas-Linares, C. Kurtsiefer, Appl. Phys. Lett. **114**, 101102 (2019).
33. H. Hong, R. Quan, X. Xiang, W. Xue, H. Quan, W. Zhao, Y. Liu, M. Cao, T. Liu, S. Zhang, R. Dong, J. Lightw. Technol., **40**, 12, 3723-3728 (2022).
34. R. Quan, H. Hong, X. Xiang, M. Cao, X. Li, B. Li, R. Dong, T. Liu, S. Zhang, New J. Phys., **26**, 09301 2(2024).
35. R. Quan, R. Dong, X. Xiang, B. Li, T. Liu, S. Zhang, Rev. Sci. Instrum. **91**, 123109 (2020).
36. C. Xiong, Christelle Monat, Alex S. Clark, Opt. Lett. **36**, 17, 3413-3415 (2011).
37. X. Xiang, B. Shi, R. Quan, Y. Liu, Z. Xia, H. Hong, T. Liu, J. Wu, J. Qiang, J. Jia, S. Zhang, R. Dong, Quantum Sci. Technol. 8, 045017 (2023).
38. C. Spiess, S. Töpfer, S. Sharma, A. Kržič, M. Cabrejo-Ponce, U. Chandrashekara, N. L. Döll, D. Rieländer, F. Steinlechner, Phys. Rev. Applied, **19**, 5, 54082 (2023).
39. J. Franson, Phys. Rev. A **45**, 3126 (1992).
40. Y. Baek, Y. W. Cho, and Y. H. Kim, Opt. Express **17**, 19241 (2009).
41. Y Liu, J Xing, Z Xia, R Quan, H Hong, T Liu, S Zhang, X Xiang, R Dong, Chin. Opt. Lett. **21**,3, 032701 (2023).
42. L. Galleani, Metrologia **45**, 6, S175–S182 (2008).
43. Y. Pang, J. E. Castro, T. J. Steiner, L. Duan, N. Tagliavacche, M. Borghi, L. Thiel, N. Lewis, J. E. Bowers, M. Liscidini, G. Moody, PRX Quantum, **6**, 010338 (2025).
44. B. Korzh, Q. Zhao, J. Allmaras, S. Frasca, T. Autry, E. Bersin, A. Beyer, R. Briggs, B. Bumble, M. Colangelo, G. Crouch, A. Dane, T. Gerrits, A. Lita, F. Marsili, G. Moody, C. Peña, E. Ramirez, J. Rezac, N. Sinclair, M. Stevens, A. Velasco, V. Verma, E. Wollman, S. Xie, D. Zhu, P. Hale, M. Spiropulu, K. Silverman, R. Mirin, S. Nam, A. Kozorezov, M. Shaw & K. Berggren, Nat. Photonics **14**, 4, 250–255 (2020).
45. J. Zhang, R. Yang, X. Li, C. Sun, Y. Liu, Y. Wei, J. Duan, Z. Xie, Y. Gong, S. Zhu, Adv. Photonics, **5**, 3, 036003 (2023).
46. J. Liu, Z. Lin, D. Liu, X. Feng, F. Liu, K. Cui, Y. Huang, W. Zhang, Quantum Sci. Technol. **9**, 015003 (2024).
47. X. Xiang, J. Liu, B. Shi, H. Hong, X. Sun, Y. Liu, R. Quan, T. Liu, S. Zhang, W. Zhang, R. Dong, Laser & Photonics Rev., in press (2025).
48. V. Giovannetti, S. Lloyd, and L. Maccone, Nat. Photon. **5**, 222–229

(2011).
49 Chen, Yuanyuan, Ling Hong, and Lixiang Chen. Front. Phys. **10,** 892519 (2022).